\documentstyle[12pt]{article}
\begin{document}
$\;\;\;\;\;\;\;\;\;\;\;\;\;\;\;\;\;\;\;\;\;\;\;\;\;\;
 \;\;\;\;\;\;\;\;\;\;\;\;\;\;\;\;\;\;\;\;\;\;\;\;\;\;
 \;\;\;\;\;\;\;\;\;\;\;\;\;\;\;\;\;\;\;\;\;\;\;\;\;\;
 \;\;\;\;\;\;\;\;\;
  176-99-HEP $
\begin{center}
\Large
{\bf Production of fast hadron leader by QCD process} \\
\vskip 0.5cm
\large
Vladimir  Lugovoi

\vskip 0.1cm
{\small \em Department of High Energy Physics, Physical Technical Institute
of Tashkent , Mavlyanov st. 2b , Tashkent , \rm 700084 \small \em Uzbekistan}
\end{center}
\vskip 0.1cm
\begin{center}
{ \em Electronic address : lugovoi@lu.silk.org }
\end{center}

\vskip 0.7cm

\begin{abstract}
An algorithm of Monte Carlo code for the production of fast hadron leader
in the proton-(anti)proton interaction by QCD process is discussed.
\end{abstract}

\normalsize

\newpage
\section{Introduction}
At present one of a few most interesting questions is theoretical
explanation of the experimental data for hevi-ion collisions.
Therefore any alternative explanations of phenomena which could be
interpreted as a signature of new physics such as quark-gluon plasma
deserve careful elaboration. In the Ref.\cite{Lug99} we place for
consideration a new hadronization process based on the cascade
decay of rotating strings into hadron states.  As it had marked in
Ref.\cite{Lug99},
long ago it have been already known fact that linearity of Regge
trajectory could be explained by supposition that the hadron is
a rigid straight rotating string\cite{Green}.
If the Regge trajectories are unlimited, it gives a chance to consider
the rotating strings having large masses too.
There is no a complete theory for decay of rotating string, and so
we have been constructing a Monte Carlo
parametrization for this process\cite{Lug99}.
In our cascade string breaking model the process of breaking of
mother string into two daugher strings is repeated.
The string rotation leads to correlation between the transverse momentum
of daughter string and the mass of mother string.   Therefore the average
transverse momentum of secondary hadrons, produced by decays of rotating
strings, grows with the mass of primary string (see Fig.5 in \cite{Lug99}
and Fig.10 in \cite{Lug98}),
and so the fluctuations of the masses of primary strings (in the model for
proton-antiproton collisions) lead to explanation of experimental $P_{T}$
distribution of secondaries up to $p_{T}= 4$ Gev at 1800 GeV for
proton-antiproton interactions (see Fig.13
in Ref.\cite{Lug98} where there are preliminary results).

    For heavy-ion reactions there is widely known theoretical prediction
that hard parton-parton interaction and formation of quark-gluon plasma
are two sources of secondary particles with high transverse momenta.
We hope that our preliminary calculations shown that the string
rotation is also
the source of high transverse momenta, and so the string rotation
can lead to effect which is the very same as vapourization of drops
of quark-gluon plasma.   Therefore this result can't be ignored
in the realistic calculations of high energy reactions.

The starting point for nucleus-nucleus interactions is nucleon-nucleon
interaction, and so we should be constructed proton-(anti)proton
Monte Carlo model for high energy interaction where, according to cosmic ray
experimental data\cite{Boris85,Ivanen,Yuld99}, for interaction at energy
$E_{Lab} > 10^{16}$ eV  there is a group of very fast hadron leaders,
which have not till now the satisfactory theoretical
explanation\footnote{In the
frame of our model a qualitative explanation for cosmic ray experimental data
about alignment of hadron leaders\cite{Boris85,Ivanen,Yuld99} can be given by
additional physics connected with the fast hadron leader production
described here and with cascade decay of rotating strings into hadron
states\cite{Lug99}. However, it requires detailed description which overstep
the limits of the theme of presented paper. Therefore the detailes
will be given elsewhere. } \cite{Ivanen,Yuld99}.

For the modelling calculations of production of the fast hadron leaders
at high-energy proton-(anti)proton interactions we suppose to use
a parton model idea that (anti)proton is a collection of quasi-free partons
which share its momentum, and that the partons carry negligible
transverse momentum.
In many models of multiparticle production (see, for example,
\cite{Capella,SjoZijl} )  in the first approximation\footnote{Only valence
quarks are considered.}  the parton structure
of proton is a color-triplet quark and an antitriplet diquark, but in the
standard parton model it is many (anti)quarks and gluons\cite{SjoZijl,Eichten84}.
Therefore in our calculations we should be taken into account that
each incoming hadron  is composite object, consisting of
constituent (anti)(di)quarks which have possibility to dissociate into
quasi-free (anti)(di)quarks and gluons (constituent partons).
This process can lead to formation of excited partons by interaction
of partons of target and projectile (for example, a gluon of target can be
"absorbed" by (anti)(di)quark of projectile).
Invariant mass of excited (anti)(di)quark can be choosed as
the characteristic of excitation.  If this excitation is sufficient
to starting of QCD cascade, it should be calculated.   If this excitation is
under threshold of QCD cascade, we can generate "decay" of excited parton into
non-excited parton and hadron leader (see Fig.1a).  Therefore for the mass
of excited parton there is a threshold which is parameter of modelling
and which is a minimal invariant mass sufficient to start of QCD cascade,
and simultaneously it may be a maximal invariant mass of excited parton
sufficient to start of hadron leader production.
However, it should be marked, there is not
any restriction to production of the fast hadron leader by excited parton
with invariant mass greater of threshold sufficient to start of QCD cascade
which has priority at high invariant mass of excited parton.
Of course, fast hadron leader can be produced also after QCD cascade
when invariant mass of cascading parton becomes less than the value of
threshold.

If (constituent) (di)quark and anti(di)quark can't be decayed, they
stretch the primary string which is decayed into secondary hadrons
according to the algorithm described in \cite{Lug99}.

\vskip 0.4cm
\section{Formation of excited partons}

An algorithm of Monte Carlo (MC) code of formation of excited partons
is described in this Section.  In our model (as in Ref.\cite{Lug95})
constituent quark of proton can dissociate into quasi-free quark and gluon
which has the part $y$ of four-momentum of constituent quark. The probability
of dissociation into interval $dy$ is formally determined by perturbative
formula (see Appendix 1 where the variable $z = 1 - y$)
\begin{equation}
dW\;\; = \;\; \alpha \frac{1+(1-y)^{2}}{y} \;\; dy \;\;\;\;\;\;\;\;  at  \;\;\;\;\;\;\;\;\; y_{min}\;\;<\;\;y\;\;<\;\;1\;\;\;\;,
\end{equation}
and integral probability is determined by the formula
\begin{equation}
W\;\; = \;\; \alpha \int_{y_{min}}^{1} \frac{1+(1-y)^{2}}{y} \;\; dy \;\;\;\;\;\;\;\;  at  \;\;\;\;\;\;\;\;\; y_{min}\;\;<\;\;y\;\;<\;\;1\;\;\;\;,
\end{equation}
where
\begin{equation}
\alpha \;\; = \;\; [ \;\; \int_{y_{1}}^{1} \frac{1+(1-y)^{2}}{y} \;\; dy \;\; ]^{-1} \;\;\;\;  ,
\end{equation}
\begin{equation}
\;\;\;\;\;\;\;\;\;\;\;\;\;\;\;\; y_{min} \;\; = \;\; y_{1}  \;\;\;\;\;\;\;\; at \;\;\;\;\;\;\;\;\;\; y_{1}\;\; > \;\; y_{0} \;\; = \;\; (\varepsilon_{0} / xE) \;\;\;\;\;\; ,
\end{equation}
\begin{equation}
\;\;\;\;y_{min} \;\; = \;\; y_{0}  \;\;\;\;\;\;\;\; at \;\;\;\;\;\;\;\;\;\;\;\; y_{1} \;\; < \;\; y_{0} \;\; < \;\; 1 \;\;\;\;\; ,
\end{equation}
\begin{equation}
y_{min} \;\; = \;\; 1  \;\;\;\;\;\;\;\;\; at \;\;\;\;\;\;\;\;\;\;\;\; 1 \;\; < \;\; y_{0} \;\; \;\; \;\;\; \;\;\;\;\; ,
\end{equation}
$\varepsilon_{0}$ and $y_{1}$ are free parameters of modelling,
$x$ is part of four-momentum which is carried by constituent (anti)(di)quark
before dissociation of it, $E$ is the energy of (anti)proton
in c.m.system of collising particles.
For example, the values can be choosed as
\begin{equation}
\varepsilon_{0} \;\; \sim \;\; m_{u} \;\;\;\;\;\;\;, \;\;\;\;\;\;\; y_{1} \;\; < \;\; 0.01 \;\;\; ,
\end{equation}
where $m_{u} = 0.34 GeV$  is the mass of constituent $u$-quark.
As one can see from eqs.(2)-(6), the probability of dissociation  grows with
energy $xE$ of constituent quark.

The parts $x$ and $x'$ of four-momentum are determined by parton distribution
function for the projectile and target.  If (according to eq.(2)) there is
the process of dissociation of constituent (anti)(di)quark into (anti)(di)quark
and gluon, the gluon part $y$ of four-momentum of constituent quark is
generated by the formula
\begin{equation}
r\;\; = \;\; \int_{y_{min}}^{y} \frac{1+(1-y)^{2}}{y} \;\; dy \;\;\;\;\; /  \;\;\;\;\;\int_{y_{min}}^{1} \frac{1+(1-y)^{2}}{y} \;\; dy \;\;\;\;\; ,
\end{equation}
where $r$ is uniformly distributed random number from the range (0,1).
By analogy with eq.(8) the part $y'$ for gluon from the target can be generated.
In this case, in eq.(4) instead $y_{0}$ the value
$y'_{0} \;\; = \;\; (\varepsilon_{0} / x'E)$ should be calculated.

Let us now to consider an example for the formation of excited parton
$\overline{q'^{*}}$ by interaction
\begin{equation}
g \;\; + \;\; \overline{q'_{c}} \;\; \rightarrow \;\; \overline{q'^{*}}  \;\;\;\;  ,
\end{equation}
i.e. a gluon of target is "absorbed" by antiquark of projectile
at proton-antiproton collision\footnote{It should be marked, the subprocess
$\;q_{c} + \overline{q'_{c}} \rightarrow
 (q+g)+(\overline{q'}+g')\rightarrow
 (q+\overline{q'})+(g+g')\rightarrow
g^{*}_{1} +g^{*}_{2}  \;$ can't be a source to production of
hadron leader.}.

Before simulation of the subprocess (9), the probability of this process
should be determined. Let $\rho$ is density of current of protons to antiprotons
and $d^{3}\rho$ is given at fixed flowers of constituents $q_{c}$ and
$\overline{q'_{c}}$ from the intervals $dx$, $dx'$, $dy$ provided that
constituent antiquark $\overline{q'_{c}}$ is not dissociated; and let
$\sigma$ is cross-section of inelastic non-diffractive
$p\overline{p}$ collisions , $\sigma^{*}_{\overline{q}'}$ is cross-section
of the  subprocess (9) , then
the probability of the subprocess (9) is given by the formula
\begin{equation}
W^{*}_{\overline{q}'} \;\; = \;\; \frac{d^{3}\rho \; \sigma^{*}_{\overline{q}'} }
{\rho \; d^{3}\sigma} \;\; = \;\; \frac{\sigma^{*}_{\overline{q}'}}{\sigma} \;\;\;\; .
\end{equation}
In the eq.(10) it was taken into account that
\begin{equation}
 \frac{d^{3}\rho}{\rho} \;\; = \;\; \frac{d^{3}\sigma}{\sigma} \;\;\;\; .
\end{equation}
According to dimension of cross-section we can choose
\begin{equation}
\sigma^{*}_{\overline{q}'} \;\; \propto \;\; \frac{1}{M^{* 2}_{\overline{q}'}} \;\;
\end{equation}
for the invariant mass
\begin{equation}
M^{*2}_{\overline{q}'} \;\; = (x^{2}y^{2}+x'^{2}) M^{2} + 2xyx'(E^{2}+P^{2}) \;\;\;
\end{equation}
which is under threshold sufficient to start of QCD cascade, or
\begin{equation}
\sigma^{*}_{\overline{q}'} \;\; \propto \;\; \frac{\alpha_{s}(M^{* 2}_{\overline{q}'})}{M^{* 2}_{\overline{q}'}} \;\;
\end{equation}
for the invariant mass sufficient to start of QCD
cascade\footnote{In ref.\cite{Lug95} there are examples of cross-sections for
the more complicated subprocesses.}.
In eqs.(12)-(14) $M^{*}_{\overline{q}'}$ is invariant mass of excited antiquark
$\overline{q'^{*}}$ from (9) and
$P$ is the momentum of (anti)proton in c.m.system of collising particles,
$\alpha_{s}(M^{* 2}_{\overline{q}'})$ is the strong coupling constant.
However, the probability value should be at least restricted.
Therefore for small invariant mass which is under threshold sufficient
to start of QCD cascade, we can choose an approximation of
probability for the process (9) as
\begin{equation}
W^{*}_{\overline{q}'} \;\; \propto \;\;\frac{\beta^{2}}{M^{*2}_{\overline{q}'} \;\; + \;\; \beta^{2} } \;\;\;\;\; ,
\end{equation}
where $\beta$ is free parameter of modelling.

The eqs.(2)-(8), (10), (13)-(15) are sufficient for the modelling of
starting-point for QCD cascade or production of fast hadron leader.

\vskip 0.4cm
\section{Fast hadron leaders}
Short distance QCD interaction leads to formation of excited (anti)(di)quark
state with short life time which is less than time sufficient for the formation
of long distance string. Therefore the process of "decay" of  excited
(anti)(di)quark with invariant mass under threshold sufficient
to start of QCD cascade , i.e. the process of fast hadron leader production
(see Fig.1a) is before formation of string.

In Fig.1a it is shown a "decay" of excited quark $q^{*}$. Therefore
this diagram is related with QCD diagram in Fig.1b  where a quark $q$ is carried
the part of four-momentum $z$ of excited quark $q^{*}$ distributed
(see Appendix 1) by the law
\begin{equation}
f(z) \;\; \propto \;\; \frac{1\;+\;z^{2}}{1\;-\;z} \;\;\;\;\;.
\end{equation}
Therefore in Fig.1a we can choose for the "decay" of excited
(anti)(di)quark $q^{*}$ the same variable $z$ and law (16)
which has peak in the point $z$=1 , i.e. hadron  $h$ is carried
the part of four-momentum $z$ of excited (anti)(di)quark $q^{*}$ near by 1 ,
i.e. this hadron $h$ is fast. In the Fig.1a slow quark $q$
stretchs the string which is decayed into secondary hadrons
according to the algorithm described in \cite{Lug99}.
Therefore in Fig.1a the hadron $h$ is fast leader.

In the frame of our model, each proton-(anti)proton interaction leads to
production at least of one fast hadron leader which carrys large part
of energy of primary particle (see eq.(16)).
This process is repeated at second interaction (in the atmosphere), and so,
in our model , it can be understood an experimental phenomena
that hadron leader has small dissipation of energy at interactions
in the atmosphere\cite{Ivanen}.

\vskip 0.4cm

{\large  \bf Appendix 1}
\vskip 0.5 cm
Let us remind of some results of QCD\cite{Halzen} which are important under
construction of our model.  We take a look at current of virtual photons
$\gamma^{*}$ to quarks $q$ (Fig.2) at fixed c.m.energy $\sqrt{s}$.
Let this current of virtual photons have uniform distribution in variable
\begin{equation}
z \;\; = \;\; \frac{Q^{2}}{s\;+\;Q^{2}} \;\;\;\;\;\;\;\; ,
\end{equation}
where $Q^{2}$ is squared four-momentum of photon taken with opposite sign.
Differential cross-section of the process given by Fig.2 is
\begin{equation}
d^{2}\sigma \; = \; \sigma_{1} \;\; \gamma_{2}(p_{T}^{2},z) \;\; dp_{T}^{2} \;\; dz
\;\;\;\;\;\;\;\;\;   at \;\;\;\;\; -t \; \ll \; s \;\;\;\; ,
\end{equation}
where
\begin{equation}
\gamma_{2} \; = \; \frac{\alpha_{s}}{2\pi} \;\;\; \frac{1}{p_{T}^{2}} \;\;\;
\frac{4}{3} \;\;\;  \frac{1\;+\;z^{2}}{1\;-\;z} \;\;\;\;\; ,
\end{equation}
$\sigma_{1}$ is cross-section of the block 1 in Fig.2 , $\alpha_{s}$ is
the strong coupling constant, $p_{T}$ is gluon transverse momentum given
relative to  momentum vector of primary quark in c.m.s. , $t$ is the squared
difference between the momenta of primary quark and gluon (see Fig.2).
It is important , at $\;\; -t \; \ll \; s \;\; $ an intermediate quark
carrys the part $z$ of four-momentum of primary quark in c.m.s.

\newpage
\begin{center}
{\large  \bf Figure Captions}
\end{center}
\vskip 0.7 cm

{\large \bf Fig.1 } $\bf a.\;$The "decay" of excited parton $q^{*}$ into non-excited
                    parton $q$ and hadron leader $h$ . $\;\;\;\bf b.\;$ QCD
                    diagram for the decay of excited quark $q^{*}$ into
                    quark $q$ and gluon $g$ .

\vskip 0.8 cm
{\large \bf Fig.2 } One in a few diagrams of the process
                    $\gamma^{*} \; q \; \rightarrow \; q \; g$ .


\newpage
\begin{thebibliography}{99}

\bibitem{Lug99} V. Lugovoi , submitted to Comp.Phys.Comm.

\bibitem{Green} M. Green, J. Schwarz, E. Witten ;  Superstring
theory,v.1,Introduction, Cambridge Univ.Press,NY,1985, p.518 .

\bibitem{Lug98} V. Lugovoi ,  hep-ph/9811486.

\bibitem{Boris85} A.S. Borisov et al., Izvestiya AN SSSR, Ser.Fiz. v.{\bf 49},
 No 7,p.1285(1985).

\bibitem{Ivanen} I.P. Ivavenko at al. Pis'ma v ZhETF, v.{\bf 5,6}, 192 (1992).

\bibitem{Yuld99} T.S.Yuldashbaev et al., Izvestiya AN, Ser.Fiz. v.{\bf 63},
No 3, p.442(1999).

\bibitem{Capella} A. Capella, U. Sukhatme, Chung I Tan, J. Tran Thanh Van,
     Phys.Lett. {\bf 81B}, 68 (1979);
     A. Capella, U. Sukhatme, J. Tran Thanh Van,
     Z.Phys.C {\bf 3}, 329 (1980) ;
     A. Capella, J. Tran Thanh Van,
     Z.Phys. {\bf C10}, 249 (1981); Phys.Lett. {\bf 114B}, 450 (1982);
     A. Capella, U. Sukhatme, C-I Tan, J. Tran Thanh Van,
      Orsay preprint LPTHE 92-38.

\bibitem{SjoZijl} T. Sjostrand and M. Zijl, Phys.Rev. {\bf 36}, 2019 (1987).

\bibitem{Eichten84} E. Eichten et al. Rewievs of Modern Phys.
     {\bf 56}, 579 (1984).

\bibitem{Lug95} V.V. Lugovoi and V.M. Chudakov, Journal of Physics of Uzbekistan,
    {\bf 6}, 8(1995).

\bibitem{Halzen}F. Halzen and A. Martin ; Quarks and leptons, John Wiley-Sons,
      New York, 1984, p.456 .

\end{thebibliography}
\end{document}